\documentclass[showpacs,aps,prb,preprint,floatfix]{revtex4-1}
\usepackage{bm}
\usepackage{graphicx}
\usepackage{epsfig}
\usepackage{amsmath,graphics,epsfig,color,verbatim}
\usepackage{bbold}
\usepackage{float} 
\usepackage{lipsum}
\makeatletter
\newcommand*{\rom}[1]{\expandafter\@slowromancap\romannumeral #1@}
\newcommand{\cRPA}{{\mathchoice{}{}{\scriptscriptstyle}{}c\!R\!P\!A}}

\setcitestyle{super}

\newcommand*{\citen}[1]{%
  \begingroup
    \romannumeral-`\x 
    \setcitestyle{numbers}%
    \cite{#1}%
  \endgroup   
}

\begin{document}
\title{ Intermolecular Coupling and Superconductivity in Chevrel Phase Compounds}
\author{Jia Chen$^{1}$, Andrew J. Millis$^{2,3}$, David R. Reichman$^1$ }
\address{$^1$Department of Chemistry, Columbia University, New York, New York 10027, USA}
\address{$^2$Department of Physics, Columbia University, New York, New York 10027, USA}
\address{$^3$Center for Computational Quantum Physics, The Flatiron Institute, New York, New York, 10010, USA}
\date{\today}

\begin{abstract}
To understand superconductivity in Chevrel phase compounds and guide the search for interesting properties in materials created with Chevrel phase molecules as building blocks, we use ab-initio methods to  study the properties of single Mo$_6$X$_8$ molecules  with $X=S$, $Se$, $Te$ as well as the bulk solid PbMo$_6$S$_8$. In bulk PbMo$_6$S$_8$, the different energy scales  from strong to weak  are: the band kinetic energy, the intra-molecular Coulomb interaction, the on-molecule  Jahn-Teller energy and the Hund's exchange coupling. The metallic state is stable with respect to Mott and polaronic insulating states. The bulk compound is characterized by a strong electron-phonon interaction with the largest coupling involving phonon modes with energies in the range from 11 meV to 17 meV and with a strong inter-molecule (Peierls) character. A two-band Eliashberg equation analysis  shows that the superconductivity is strong-coupling, with different gaps on the two Fermi surface sheets. A Bergman-Rainer  analysis of the functioanl derivative of the transition temperature with respect to the electron-phonon coupling reveals that the Peierls modes provide the most important contribution to the superconductivity. This work illustrates the importance of inter-molecular coupling for collective phenomena in molecular solids.
\end{abstract}

\maketitle

\section{Introduction}\label{intro}
Synthetic chemists are now able to assemble  molecular clusters into crystal structures with atomic precision\cite{Roy2013}, making the search for collective and emergent properties in those super-atomic solids a timely and important  topic. The notion of bootstrapping interesting molecular properties and strong molecular interactions into important bulk properties is an important theme in the field. For example,  the relatively high transition temperature superconductivity in some members of the alkali-doped fullerenes is believed to arise from  intra-molecular  vibrational modes \cite{Varma1991, Gunnarsson1997} whereas in other alkali-doped fullerenes it is argued \cite{Takabayashi2009}  to arise from   intra-molecular electron-electron interactions. The recent discovery of superconductivity in endohedral gallide clusters  also exemplifies the rich set of possibilities provided by molecular solids. \cite{Xie2015}

Binary and ternary molybdenum chalcogenides, also known as Chevrel phase compounds\cite{Chevrel1971} are of great interest in this context. Their chemical formula is  M$_m$Mo$_6$X$_8$, where M is a metal element and  X=S, Se, Te. The bulk compound can be viewed as a molecular crystal of Mo$_6$X$_8$ units on the sites of a  rhombohedral lattice, with the metal ions  M in interstitial sites. The materials have been of sustained interest to both physicists and chemists because they can be superconducting with transition temperatures as high as $15$ K (PbMo$_6$S$_8$)\cite{Matthias1972} and a high upper critical field.\cite{Matthias1972, Odermatt1974, Fischer1973} Despite some hints at unconventional superconductivity, \cite{Uemura1991, Dubois2007}  it is generally accepted that the electron-phonon interaction provides the pairing mechanism.\cite{Culetto1978, Petrovic2011} Chevrel compounds  have also been proposed as promising multivalent cathode materials in Mg batteries.\cite{Aurbach2000} Recent experimental efforts have been directed at synthesis of lower-dimensional Chevrel phase compounds.\cite{Zhong2018}

Since Chevrel phase compounds are built  of Mo$_6$X$_8$ molecular clusters, it is natural to approach the its physics via a model of relatively weakly coupled clusters.\cite{Bullett1977, Mattheiss1977, Andersen1987} The role of the intra and inter-cluster vibrational modes\cite{Pobell1982} in the superconductivity needs to be established.

In this paper we analyze PbMo$_6$S$_8$ as a model system to gain insight into the role of intra and inter-site interactions in molecular crystals and into the specifics of superconductivity in the Chevrel marterials. To approach this system, we first calculate properties of isolated Mo$_6$X$_8$ molecules and use the results to derive and parametrize effective Hamiltonians including electron-electron and electron-phonon couplings. We study the bulk properties of PbMo$_6$S$_8$, calculating electron and  phonon band structures, the electron-phonon coupling and the intra-molecular Coulomb interaction. Migdal-Eliashberg theory is then used to calculate the phonon renormalization of the bands and the superconducting gap functions and transition temperatures. Our key result is that the picture of intra-molecular interactions combined with weak constant electronic hopping between molecular units is not an adequate description of the bulk compounds. Inter-molecule effects, most notably phonons that simply do not exist in the single molecule case except as a translation or a rotation of model, play a crucial role in setting the electronic properties including superconductivity while the intra-molecular couplings have significantly weaker effects. Screening of the intra-molecular Coulomb interaction is of significant importance important in Chevrel phase compounds.

This manuscript is organized as follows. In section~\ref{isolated}, we consider isolated Mo$_6$X$_8$ molecules, identifying the important low energy degrees of freedom and interactions within the building blocks of the solids. Section~\ref{property} and \ref{phonons} discuss electron and phonon band structures, Hubbard U, Hund's exchange J and electron-phonon interaction in bulk PbMo$_6$S$_8$. In section~\ref{CEP}, we present the consequences of the electron-phonon interaction and diagnose which phonons are most important for superconductivity. Section~\ref{conclusion} is a conclusion.

\section{Molecular Properties \label{isolated}}

Molecular solids such as the Chevrel phase materials are composed of molecular building blocks (Mo$_6$X$_8$ in the present case) held together with other elements (metal ions such as Pb, in the present case). The first step in understanding the properties of molecular solids is to determine the relevant orbitals of the building blocks, and the electron-electron and electron-phonon interactions relevant to these orbitals. To obtain this information we study properties of isolated netural and charged Mo$_6$X$_8$ molecules using Density Functional Theory (DFT) methods  with the PW91 generalized gradient approximation exchange-correlation functional\cite{Perdew1992} as implemented in the {\it NWChem} package.\cite{NWChem} The basis set for molybdenum, selenium, tellurium is LANL2DZ,\cite{Hay1985} and for sulfur is 6-31G**.\cite{Petersson1988, Petersson1991}  

\begin{figure}
\includegraphics[width=1.0\columnwidth,clip]{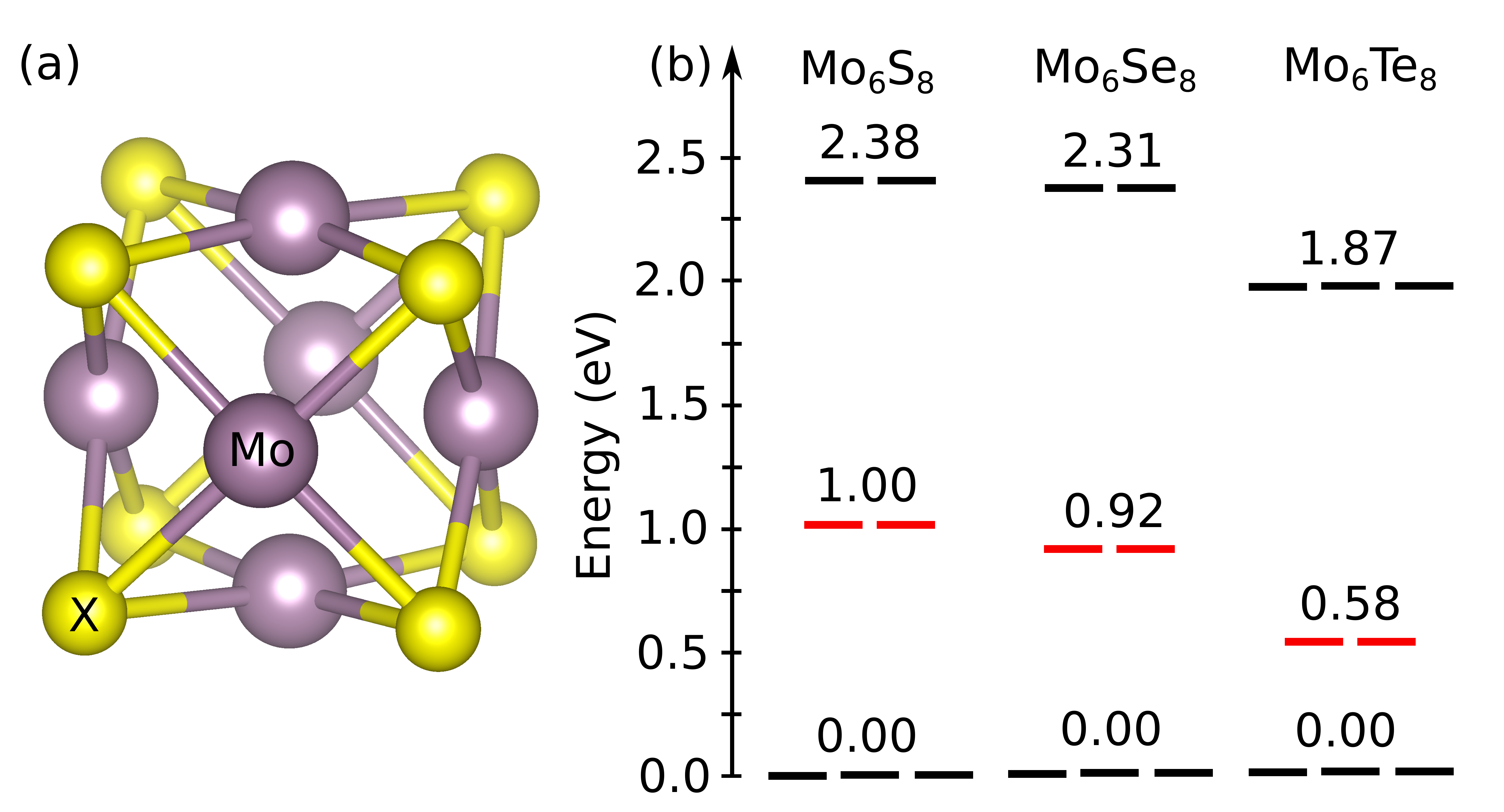}
\caption{Panel(a): Structure of Mo$_6$X$_8$ molecule; Panel (b): relative HOMO, first and second LUMO levels of neutral Mo$_6$S$_8$, Mo$_6$Se$_8$ and Mo$_6$Te$_8$. The HOMO levels of all three are set to 0.} \label{molecule_level}
\end{figure}

\begin{figure}
\includegraphics[width=1.0\columnwidth,clip]{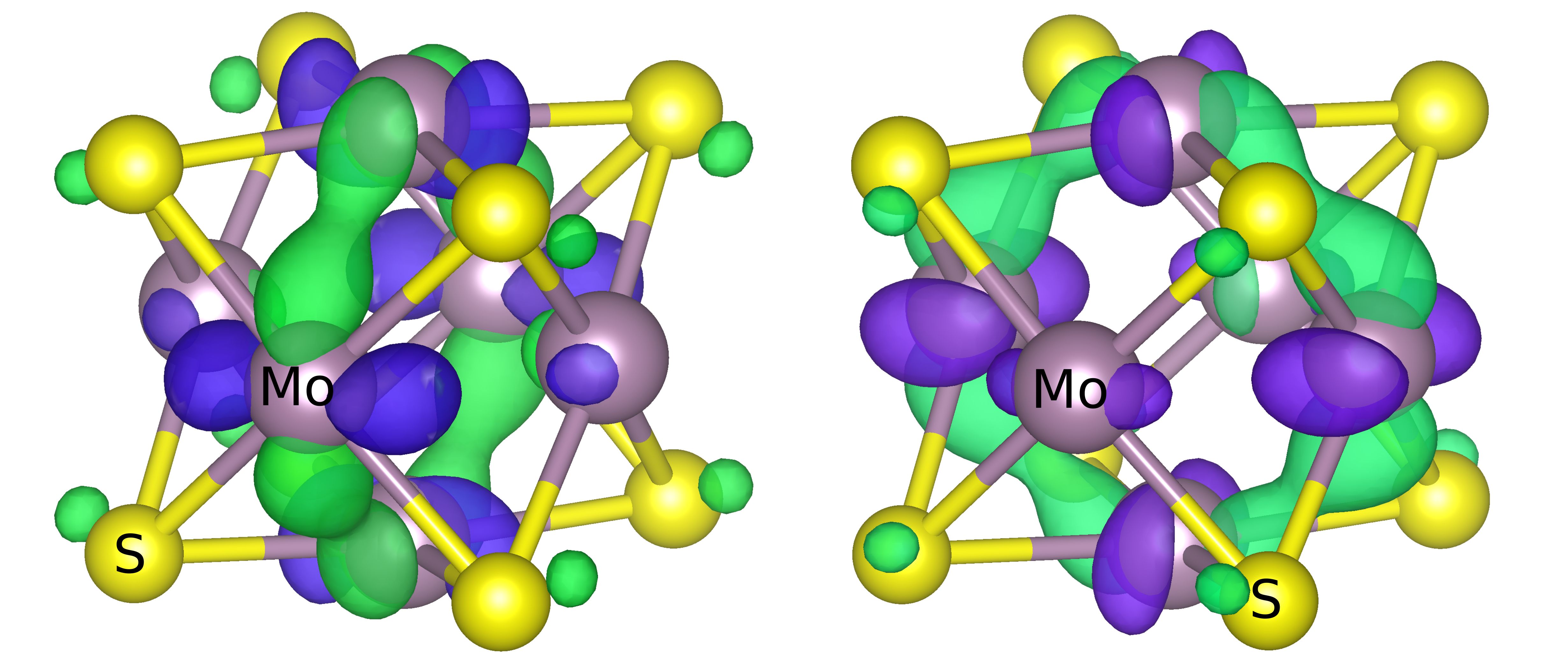}
\caption{Two-fold degenerate LUMO orbitals of neutral Mo$_6$S$_8$. The value for iso-surface in the plot was chosen to be 0.02.} \label{LUMO}
\end{figure}

Neutral Mo$_6$X$_8$ molecules (shown in panel (a) of Fig.~\ref{molecule_level}) have the symmetry of the {\it O$_h$} point group. Panel (b) of Fig.~\ref{molecule_level} shows that the highest occupied molecular orbitals (HOMO) are  three-fold degenerate while the  lowest unoccupied molecular orbitals (LUMO) are  two-fold degenerate and transform according to the $E_g$ representation of $O_h$. We focus on the LUMO doublet here because in the bulk solids of interest the $M$ ions transfer electrons to the Mo$_6$X$_8$ clusters, so the Fermi level lies in bands derived from these orbitals. Energetically, the {\it E$_g$} orbitals are separated from other molecular orbitals by 1.0 eV in Mo$_6$S$_8$; this separation becomes smaller for Mo$_6$Se$_8$ and Mo$_6$Te$_8$. Plots of the {\it E$_g$} orbitals are shown in Fig.~\ref{LUMO}: each of those two orbitals approximately consists of  {\it d$_{x^2-y^2}$} orbitals arising from four coplanar Mo ions.

We can estimate the intra-molecular electron-electron interaction $U$ of isolated Mo$_6$X$_8^{2-}$ from the charging energy:  $U=E_{Mo_6X_8^{3-}} +E_{Mo_6X_8^{1-}} - 2 E_{Mo_6X_8^{2-}} $ and Hund's exchange J from the energy difference between singlet and triplet: $2J =E^{singlet}_{Mo_6X_8^{2-}} - E^{triplet}_{Mo_6X_8^{2-}} $ From Table.~\ref{JT}, we can see  $U \approx 3.5$ eV for Mo$_6$X$_8$ molecules. $J\approx 100$ meV for all three molecules, and is just large enough to overcome the Jahn-Teller electron phonon coupling in Mo$_6$X$_8^{2-}$ molecules as discussed below. 

We now turn to the electron-phonon coupling, focussing on those modes that couple linearly  to the LUMO orbitals. Phonons couple to electron bilinears; the electrons transform as the $E_g$ representation of $O_h$ and the direct product of two {\it E$_g$} representations of the $O_h$ group can be reduced as  $E_g \times E_g = a_{1g} + a_{2g} + e_g$, so we need to consider only vibrational modes belonging to the {\it a$_{1g}$, a$_{2g}$} and {\it e$_g$} representations. The {\it e$_g$} mode is Jahn-Teller active, which means it can lift the degeneracy and lower the symmetry of the molecule. The phonon frequencies and normal mode vectors are computed by diagonalizing Hessian matrix, leading to a phonon plus electron-phonon Hamiltion which we write representing the phonons in a first quantized form using a normalized phonon operator $Q$. For $A$-symmetry (scalar) phonons we have
\begin{align}
\mathcal{H}(Q_\alpha)=\frac{\hbar \omega_{\alpha}}{2}(-\frac{\partial^2}{\partial Q_{\alpha}^2} + Q_{\alpha}^2)+g_\alpha Q_\alpha n_{el} ,
\label{Hepscalar}
\end{align}
where $n_{el}$ is the number of electrons in the LUMO states, and $\alpha$ labels phonon modes.

For $e$-symmetry (doublet) phonon modes we represent the mode as a two component vector $\vec{Q}=(Q^x,Q^z)$ and write
\begin{align}
\mathcal{H}(\vec{Q}_\alpha)=\frac{\hbar \omega_{\alpha}}{2}(-\frac{\partial^2}{\partial \vec{Q}_{\alpha}^2} +\left| \vec{Q}_{\alpha}\right|^2)+g_\alpha \vec{Q}_\alpha \cdot \sum_{ab\sigma}d^\dagger_{a\sigma}\vec{\tau}^{ab}d_{b\sigma},
\label{Hepdoublet}
\end{align}
where $\tau$ is a Pauli matrix and $a,b$ label the two states of the electronic $E_g$ doublet. 

\begin{figure}
\includegraphics[width=1.0\columnwidth,clip]{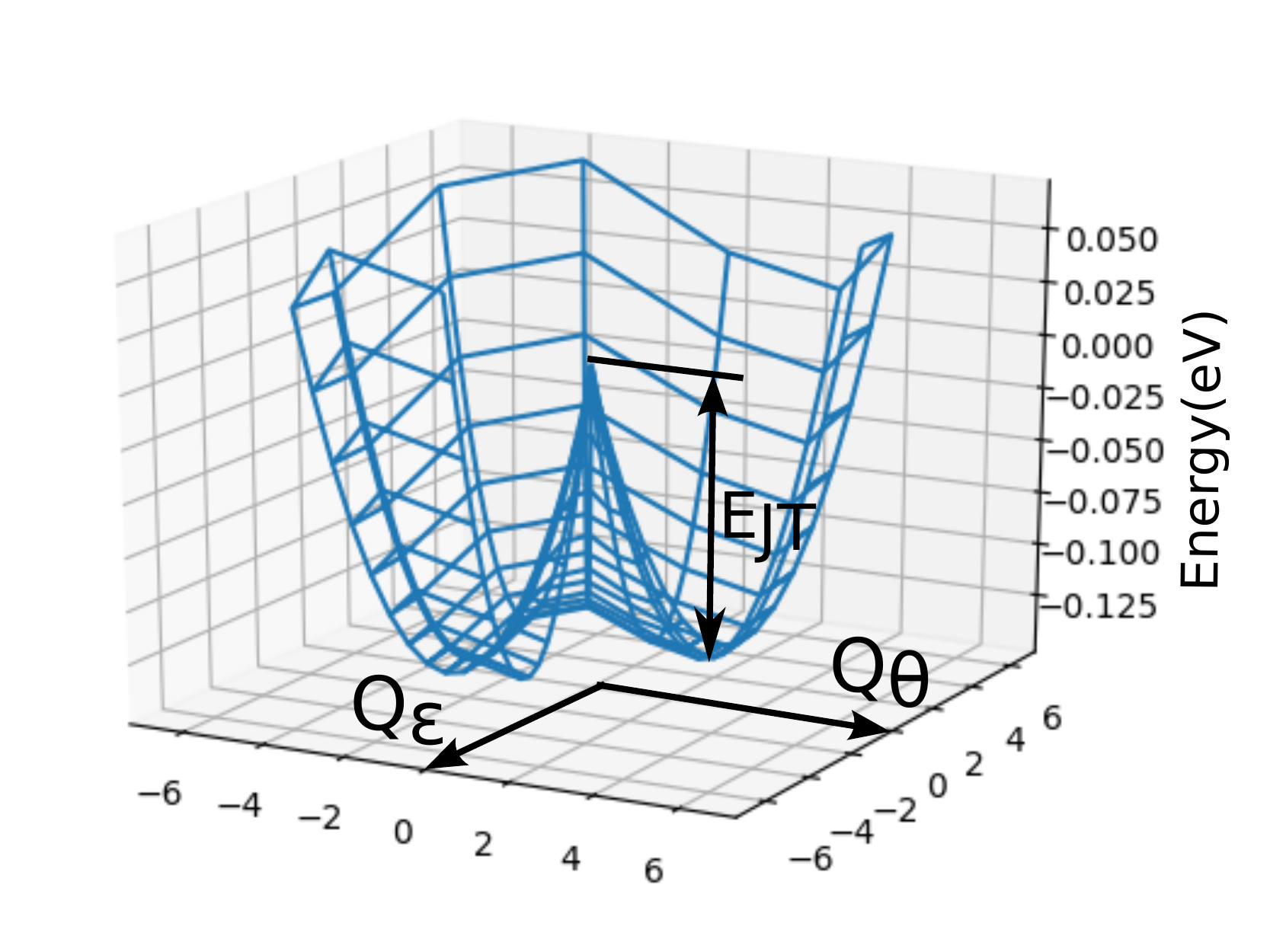}
\caption{Adiabatic potential energy calculated for the vibrational mode of Mo$_6$S$_8$ at 32.8 meV with the occupation of two electrons ($n=2$).} \label{APES}
\end{figure}

The adiabatic potential energy surface (APE) for phonon mode $\alpha$ is defined as the ground state eigenvalue of Eq.~\ref{Hepscalar},\ref{Hepdoublet} with the kinetic energy ($\partial_{Q_\alpha}$) terms neglected. The difference between the value at the minimum and the value at $Q=0$ defines the phonon stabilization energy
\begin{equation}
\omega_{eff,\alpha}=\frac{g_\alpha^2\rho_{el}^2}{2\omega_\alpha}.
\label{stabilization}
\end{equation}

Here $\rho_{el}$ is the LUMO occupancy for the $A$-symmetry modes and is  the maximal orbital disproportionation ($\rho=1$ for $n=1,3$ and $\rho=2$ for $n=2$)  in the $E$ (Jahn-Teller) case. The coupling constants $g_\alpha$ are determined from the calculated APES.

For $A$-symmetry phonons the APE is a parabola with minimum at $Q_\alpha=-\frac{g_\alpha n_{el}}{\omega_\alpha}$. We find two $A$ modes, with frequencies of $41.6$ and $50.4$ meV. The associated stabilization energies are $0.2$ and $7$ meV, respectively, too small to be of relevance to the issues discussed here. We neglect the $A$ symmetry phonons henceforth.

\begin{table}
\caption{ Quadratic frequency, linear coupling energy, and the Jahn-Teller stabilization energy for occupation number $n= 1, 2, 3$. For the neutral molecule, $\omega_1=32.8$meV and $\omega_2=29.9$ meV. For occupation $n=4$, the Jahn-Teller effect is no longer active, but mode softening is still visible.}\label{eeaa}
\begin{tabular}{l | c c|c c|c c}
\hline
 Occupation & $n=1$ & $n=1$ & $n=2$ & $n=2$ & $n=3$ & $n=3$  \\
modes & $\omega_1$  & $\omega_2$ & $\omega_1 $ &$\omega_2$ & $\omega_1$ &$\omega_2$ \\
 \hline
$\omega$ (meV) & 31.8   & 28.4  & 31.0 & 27.8 & 28.5 & 27.3 \\
g (meV)  & 49.4 & 27.0 & 46.0 & 23.8 & 38.8 & 18.8 \\
k= g/$\omega$ & 1.55 & 0.95 & 1.48 & 0.86 & 1.36 & 0.69  \\ 
E$_{JT}$ (meV) & 38.4 & 12.8 & 136.2 & 40.8 & 26.4 & 6.5 \\
 \hline
\end{tabular}
\end{table}

For the $E$ (doublet) phonons the APES has the familiar ``mexican hat'' form shown for one of the phonons in Fig.~\ref{APES}. At the quadratic level considered here the theory has the full $O(2)$ symmetry in the phonon modes, so  energy is a function only of $\rho=\left|\vec{Q}\right|$. Higher order terms in $Q$ lift the degeneracy leading to three degenerate minima (visible on close inspection in  Fig.~\ref{APES}) as required by the $O_h$ symmetry.  We find two $E$-symmetry modes; their frequencies, linear coupling parameters, and stabilization energies as function of occupations of LUMO states are listed in Table~\ref{eeaa} for Mo$_6$S$_8$. The coupling of the mode at high frequency is much larger than that of the mode at low frequency.  As the occupation number increases, the vibrational modes becomes slightly softer and the linear coupling parameter $g$ becomes weaker. The total stabilization energy is the sum of the stabilization energies of the two modes and is shown in Table~\ref{JT} for the three different choices of calcogen ions. As the chalcogenide elements become heavier, the Jahn-Teller stabilization energy decreases significantly, which correlates with the manner in which size and flexibility of the molecules change with chalcogenide element.  

In the isolated singly charged molecule Mo$_6$X$_8^{1-}$, the Jahn-Teller effect ($E_{JT} \approx 50$ meV)  is unopposed and we expect the molecule to distort away from a cubic shape. For the doubly charged Mo$_6$X$_8^{2-}$, the Jahn-Teller energy is about four times  as large as it is for the singly charged case, however, the distortion energy is reduced by the Hund's exchange J ($\approx 100$ meV). Our calculation indicates a spin triplet ground state for Mo$_6$X$_8^{2-}$, but the energy difference is small enough that this conclusion should be treated as preliminary. The large value of the on-site Coulomb interaction, which is much greater than the $n=2$ Jahn-Teller stability energy, implies that an ensemble of singly charged molecules will not disproportionate into bipolarons. 

The Jahn-Teller stabilization energy is a useful measure for comparing the relative strengths of the Jahn-Teller effects across different material families. The stabilization energies $186$ meV  we find for Mo$_6$S$_8$ at $n=2$ is smaller than the $500$ meV found in LaMnO$_3$ \cite{Millis1996} or the 215 meV and 341 meV found for LiMnO$_2$ and LiCuO$_2$. \cite{Marianetti2001} 

\begin{table}[htbp]
\caption{ Total Jahn-Teller stabilization energy, charging energy and Hund's exchange for Mo$_6$X$_8^{2-}$.}\label{JT}
\begin{tabular}{l | c | c  |c }
\hline
  & Mo$_6$S$_8^{2-}$ & Mo$_6$Se$_8^{2-}$ & Mo$_6$Te$_8^{2-}$\\
 \hline
E$_{JT}$ (meV) & 186.0   & 144.6  & 86.1 \\
J (meV)  & 105.1 & 99.6 & 90.4 \\
U (eV) & 3.7 & 3.5 & 3.4 \\
\hline
\end{tabular}
\end{table}

\section{Bulk Compound: Electronic  Properties}\label{property}
\subsection{Electronic Band Structure\label{bs}}

We next study the electronic structure of PbMo$_6$S$_8$ solid via  DFT calculations with PBE as exchange-correlation functional,\cite{Perdew1996} as implemented in the {\it Quantum Espresso} package.\cite{QE} Unless otherwise noted the structures are fully relaxed both in terms of atomic positions and lattice constants. Norm-conserving separable dual-space Gaussian pseudopotentials \cite{Hartwigsen1998} were used for all elements. The kinetic energy cutoff for wavefunctions is 80 Rydberg and the convergence threshold for force is 1.0$\times 10^{-4}$ Hartree/Bohr.      

\begin{figure}[b]
\includegraphics[width=1.0\columnwidth,clip]{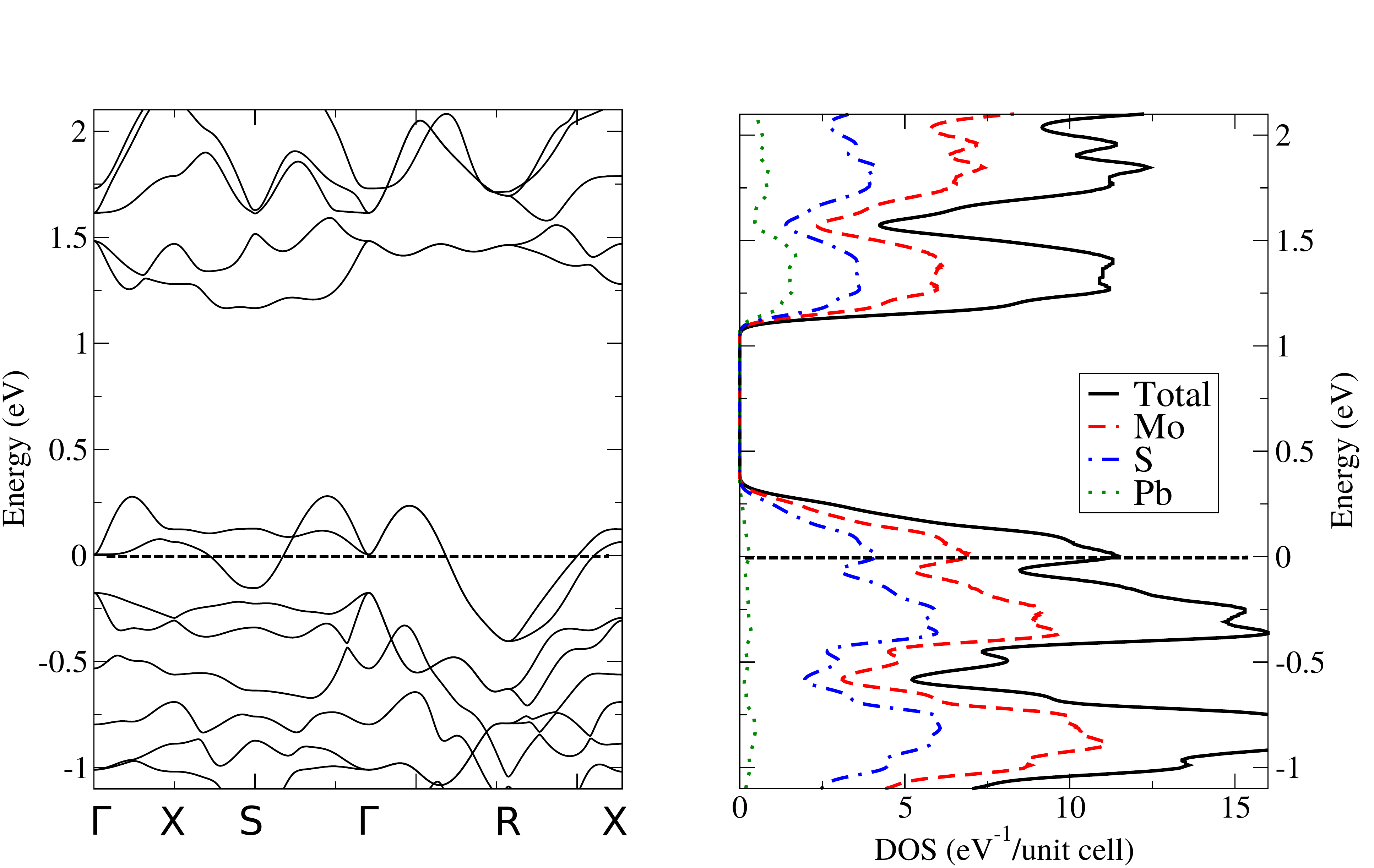}
\caption{Band structure (left panel) and density of states (right panel) of PbMo$_6$S$_8$. Since the unit cell is very close to be orthorhombic, we used high symmetry points of simple orthorhombic lattice in band structure. The Fermi level is set to 0.} \label{BS_DOS}
\end{figure}

The left panel of Fig.~\ref{BS_DOS} shows the band structure and density of states (DOS) of PbMo$_6$S$_8$. The relaxed lattice constant and bond angle are $ a = 6.55~\AA$ and  $\alpha = 89.12^{\circ}$, in good agreement with experimental values of $ a = 6.55~\AA$ and $\alpha = 89.33^{\circ} $.\cite{MAREZIO1973} Consistent with the single-molecule results, we see that only two bands cross the Fermi level; these are derived from the $E_g$ states discussed above. The band width of the two $E_g$-derived bands is $W \approx 0.7$ eV. Near the R-point the energy of the $E_g$ bands is lower than that of the other filled bands, but there are no band crossings, thus no band entanglement, enabling a straightforward Wannier analysis of the two conduction bands. 

The right panel shows the total density of states and its projection onto the component atoms. The dominant contribution to the near Fermi surface states is from Mo orbitals, with some contribution from S and negligible contribution from Pb.  For the DOS around the Fermi level, Mo contributes more than S and the contribution of Pb is negligible. The Fermi level is at a local and sharply peaked maximum in the density of states, consistent with previous arguments by Andersen and co-workers  \citep{Andersen1987} based on the pressure dependence of the superconducting transition temperature.  The total DOS at Fermi level is $N_{BS}=10.8$/(eV unit-cell), about a factor of 4 smaller than the experimental value  $N_{\gamma} = 44.4$/(eV unit-cell). (See Ref. \citen{Fischer1978} and references therein)  The dominant source of the difference is the electron-phonon coupling, as we will show below.

\begin{figure}
\includegraphics[width=1.0\columnwidth,clip]{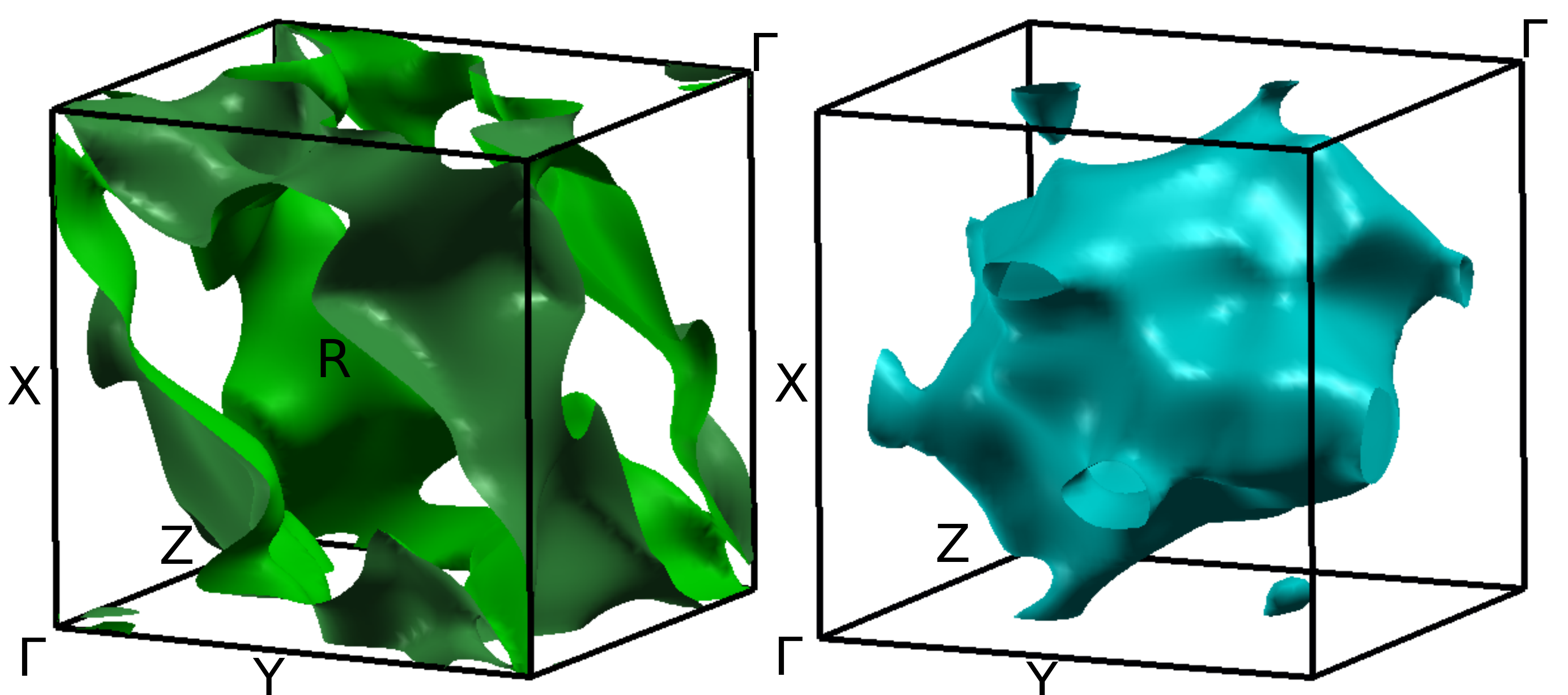}
\caption{Depiction of the Fermi surface of PbMo$_6$S$_8$ formed by the lower band (left panel) and the higher band (right panel) in the primitive cell of the reciprocal lattice } \label{FS_3D}
\end{figure}

\begin{figure}
\includegraphics[width=1.0\columnwidth,clip]{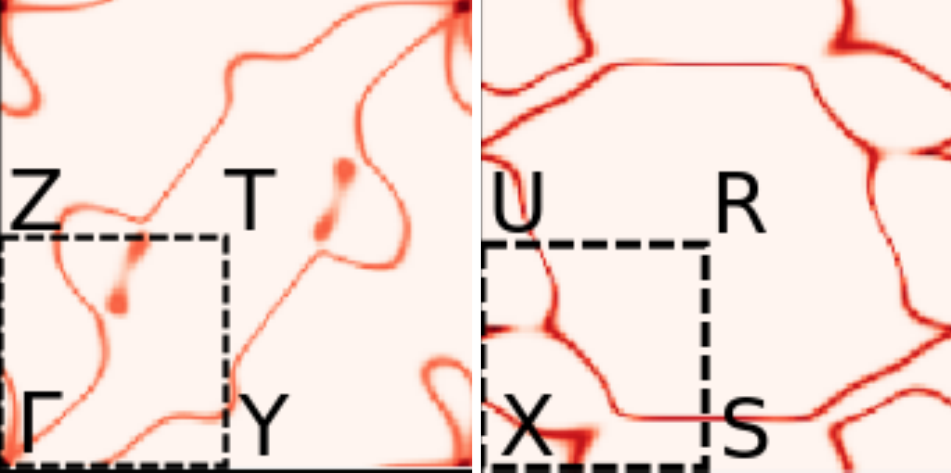}
\caption{Density of states at the Fermi level of PbMo$_6$S$_8$ in the XY planes through the $\Gamma$ point (left panel) , through the R point (right panel).} \label{FS_2D}
\end{figure}

Using the Wannier90 \cite{MOSTOFI2014} implementation of the  maximally-localised Wannier function method \cite{Marzari1997}   we studied the two  $E_g$ bands around Fermi level in some detail. From our calculation, the total occupation of these two bands is 2.0, which is consistent with a scenario in which each Pb transfers two electrons to a Mo$_6$S$_8$ cluster. The occupations of the lower and higher bands are 1.41 and 0.59, respectively. Two Fermi surfaces formed by the lower and higher bands are shown in Fig.~\ref{FS_3D}. For PbMo$_6$S$_8$, two sheets of Fermi surfaces can be found, but they are not always well separated. This has implications for superconducting order parameters, as we will discuss in section~\ref{SC}. The Fermi surface associated with the lower band centered around $\Gamma$ point is hole-like and the Fermi surface associated with the higher band is electron-like. Two dimensional cuts of the Fermi surfaces are shown in Fig.~\ref{FS_2D}. On the XY plane through the R point (right panel), two  separated Fermi surfaces are clear with the larger one as the electron pocket. It should be noted that two Fermi sheets touch each other at some places in Brillouin zone. For example, at the $\Gamma$ point, two bands can be found at Fermi level.

\subsection{Electron-electron interactions \label{ee}}

We performed constrained random phase approximation (cRPA) calculations of the effective interactions between electrons in the two frontier bands, following the approach developed by Aryasetiawan et al. \cite{Aryasetiawan2006} The polarization matrix in reciprocal space was calculated in the random phase approximation as implemented in  the {\it BerkeleyGW} package \cite{DESLIPPE2012} with a $2 \times 2 \times 2$ k-point mesh, 100 unoccupied states and kinetic energy cutoff of 5 Ry for the polarization matrix. The result is divided into contributions between states in the low energy sector ($P^{le}$), and processes involving transitions in at least one other band ($P^r$) as
\begin{equation}
P^{tot}_{\textbf{GG}'}(\textbf{q}) = P^{le}_{\textbf{GG}'}(\textbf{q}) + P^{r}_{\textbf{GG}'}(\textbf{q}).
\end{equation}
A dielectric matrix representing screening by the other degrees of freedom is constructed from $P^r$ as
\begin{equation}
\epsilon_{\textbf{GG}'}(\textbf{q}) = \delta_{\textbf{GG}'} - \nu_{\textbf{GG}'}(\textbf{q})P^{r}_{\textbf{GG}'}(\textbf{q}), 
\end{equation}
and the partially screened interaction is defined as
\begin{equation}
W(\textbf{r},\textbf{r}') = \frac{4\pi}{\Omega} \sum_{\textbf{qGG}'}\nu_{\textbf{GG}'}(\textbf{q})e^{i\textbf{(q+G)} \cdot \textbf{r}} \epsilon_{\textbf{GG}'}^{-1}(\textbf{q})e^{i(\textbf{q+G}')\cdot\textbf{r}'},
\end{equation}
where $\nu_{\textbf{GG}'}(\textbf{q})$ is the bare Coulomb interaction and $\Omega$ is the volume of the unit cell. 

The effective on-molecule interactions, namely the Hubbard U and Hund's exchange coupling J, are obtained by projecting $W$ onto the two $E_g$ orbitals of the isolated molecule:  
\begin{equation} 
U_{nm} = \int \int d\textbf{r} d\textbf{r}' \phi_{n}^*(\textbf{r}) \phi_{n}(\textbf{r}) W(\textbf{r},\textbf{r}')\phi_{m}^*(\textbf{r}') \phi_{m}(\textbf{r}'),
\end{equation}
\begin{equation} 
J_{nm} = \int \int d\textbf{r} d\textbf{r}' \phi_{n}^*(\textbf{r}) \phi_{m}(\textbf{r}) W(\textbf{r},\textbf{r}')\phi_{n}^*(\textbf{r}') \phi_{m}(\textbf{r}').
\end{equation}

The bare and screened local electron-electron interactions are listed in Table.~\ref{cRPA}. The bare interactions are larger than the charging energies reported in section \ref{isolated} because the isolated molecule calculations include relaxation of other electronic degrees of freedom (on-molecule screening).  We find that the screening is almost complete; the screened interactions are factors of $\sim 20$ less than the bare interactions, in contrast to other other molecular materials including $\kappa$-ET organic,\cite{Nakamura2009} alkali-doped C$_{60}$ and  aromatic compounds.\cite{Nomura2012} The strong reduction of the interaction can be understood in terms of the very large dielectric constant arising from the rest of the bands, $ \epsilon_{\cRPA}=\lim_{\textbf{G}+\textbf{q} \to 0} 1.0 / {\epsilon_{\textbf{GG}}^{\cRPA}}^{-1} ( \textbf{q} ) = 24.0$.  We also observe that, in contrast to the simple perovskite transition metal oxides even the Hunds coupling is significantly renormalized, consistent with reported results for organic molecular materials.\cite{Nomura2012}

Given the band width $W\approx 0.7$ eV found in band structure calculations, the interaction strengths we find confirm that PbMo$_6$S$_8$ is far from the Mott transition regime and that local correlation effects may be neglected. We may simply consider the materials to be metals with essentially weak electronic correlations. 

\begin{table}
\caption{ The values of  the bare and screened local interactions in eV.}\label{cRPA}
\begin{tabular}{c  c  c  c  c c}
\hline
  $U_{bare}$ & $U_{\cRPA}$ & $U'_{bare}$  & $U'_{\cRPA}$ & $J_{bare}$ & $J_{\cRPA}$  \\
 \hline
5.36 & 0.28 & 5.07   & 0.18 & 0.139  & 0.044 \\ 
\hline
\end{tabular}
\end{table}

\section{Phonon band structure and electron-phonon coupling \label{phonons}}

\subsection{Phonon Band Structure}

\begin{figure}
\includegraphics[width=1.0\columnwidth,clip]{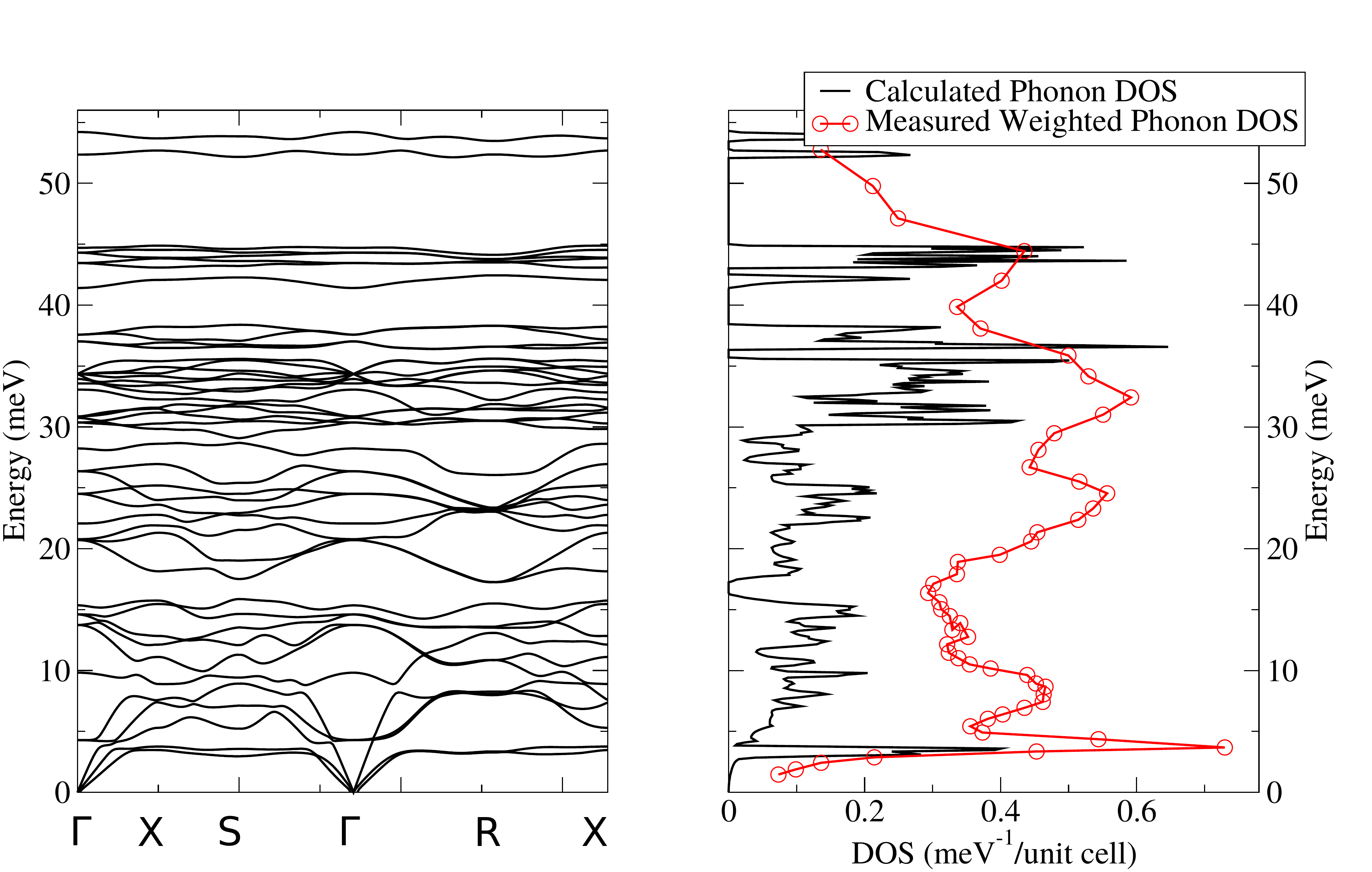}
\caption{Calculated phonon band structure (left panel), and calculated phonon density of states and measured neutron weighted phonon density of state (right panel), from Ref.~\protect\citen{Schweiss1976}.} \label{phonon}
\end{figure}

Starting from the fully relaxed electronic structures presented in the previous section we used density functional perturbation theory  (DFPT)\citep{Baroni2001} to calculate the phonon band structure and density of states shown in Fig~\ref{phonon}. The calculated phonon density of states agrees reasonably well with the density of states inferred from neutron scattering experiments.\cite{Schweiss1976} Both calculation and experiment show a sharp peak at about 4 meV, and two gaps around 17 meV and 40 meV. 

We have calculated the normal modes and find that the sharp peak in the phonon DOS at 4 meV arises from  two modes with large Pb displacements (these modes also contribute to the very large dielectric constant), in agreement with the experimental observation that the peak is absent in Chevrel phase compounds without Pb ions. \cite{Schweiss1976, Bader1976} Previous work had suggested that the minimum in the DOS at 17 meV marked the separation between internal (on-molecule) and external (intermolecular) vibrations.\cite{Bader1976_prl,Schweiss1976} We find 2 internal modes below 17 meV, which suggests hybridization between internal and external modes is present  below 17 meV,  similarly to the result found with Born-von K\'{a}rm\'{a}n lattice dynamics calculations with Lennard-Jones potentials.\cite{Bader1978} 

\subsection{Electron-Phonon Coupling} \label{e_ph}
We have used DFPT to  calculate the  matrix elements $g^{\upsilon}_{ij}(\textbf{k},\textbf{p})$  describing the scattering of an electron at momentum $\textbf{p}$  in band $j$ to momentum $\textbf{k}$ in band $i$ by emission or absorption of a phonon mode $\upsilon$ at momentum $k-p$.  The calculations were performed on  a $4 \times 4 \times 4$ grid in the Brillouin zone and then interpolated onto  fine grids via electron and phonon Wannier functions following Refs. \citen{Giustino2007, Giustino2017} as implemented in the {\it EPW} code.\cite{Ponce2016}  The fine electron  grid is  $32 \times 32 \times 32$ and the fine phonon grid is $16 \times 16 \times 16$. Convergence of the electron-phonon coupling constant with respect to coarse and fine grids sizes has been verified.

From the matrix elements we calculate the band-resolved electron-phonon coupling function $\alpha^2F$ as
 
\begin{align}\label{a2f_eq}
 \alpha^2F_{ij}(\nu) = \frac{1}{N_i(0)} \sum_{\textbf{k},\textbf{p},\upsilon} |g^{\upsilon}_{ij}(\textbf{k}, \textbf{p})|^2 
 \delta(\epsilon^i_\textbf{k}) \delta(\epsilon^j_{\textbf{k}-\textbf{p}}) \delta(\nu-\omega_{\textbf{p}}^\upsilon) ,
\end{align}
and the  band-resolved total energy-phonon coupling constant as
\begin{align}\label{lambda}
\lambda_{ij} = 2 \int_{0}^{\infty} \frac{\alpha^2F_{ij}(\nu)}{\nu} d\nu . 
\end{align}

\begin{figure}
\includegraphics[width=1.0\columnwidth,clip]{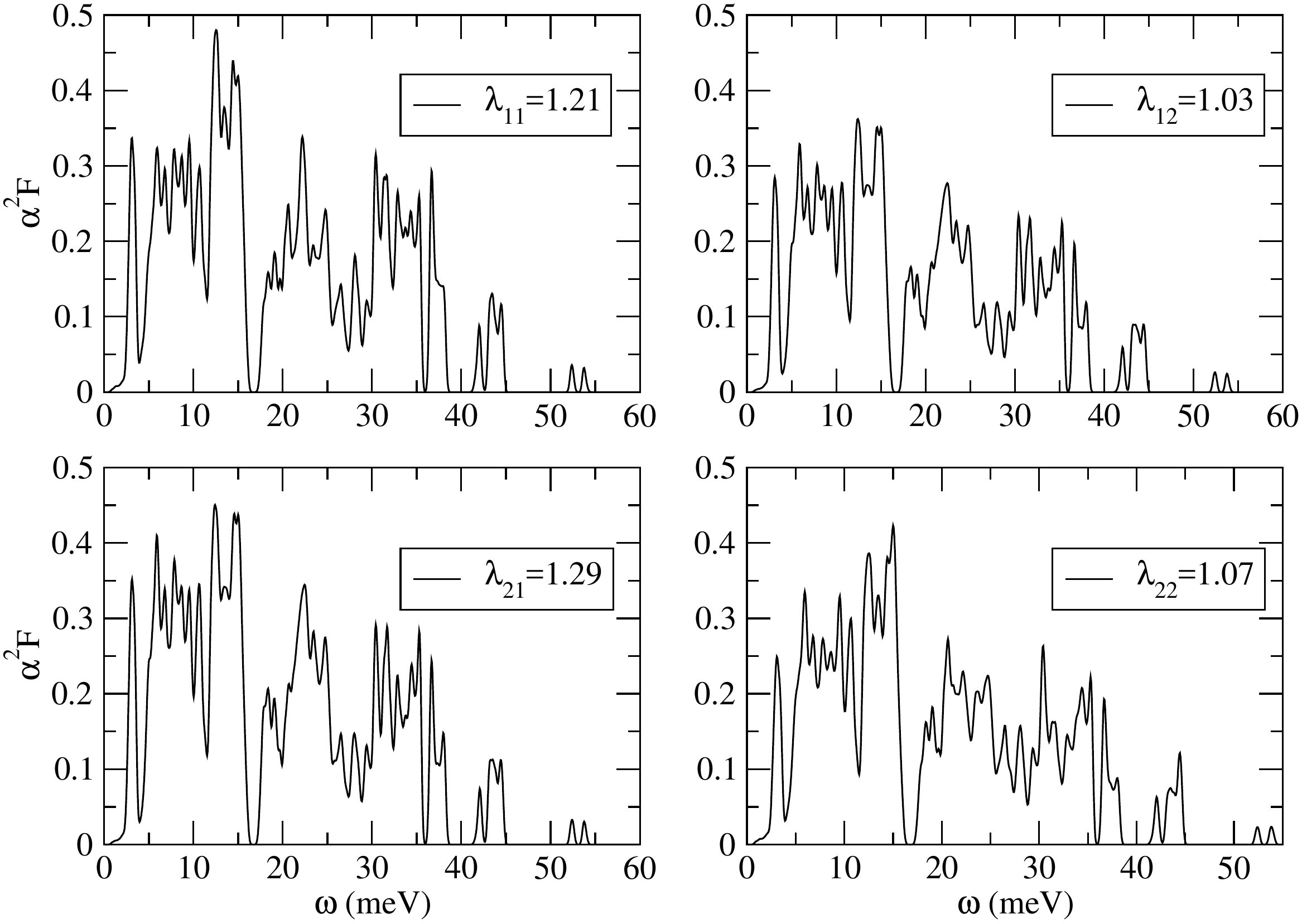}
\caption{Band-resolved electron-phonon interaction functions $\alpha^2F$ from Eq. \ref{a2f_eq} and electron-phonon coupling constants, from Eq. \ref{lambda}.} \label{a2f}
\end{figure}

Band-resolved electron-phonon spectral functions and coupling constants are shown in Fig.~\ref{a2f}. The four $\alpha F_{ij}$ have similar structures, and give similar coupling constants. This is very different from the two-band superconductor MgB$_2$, for which  intra-band coupling is much stronger than inter-band coupling.\cite{Golubov2002} We believe the difference arises because in MgB$_2$ the two bands arise from physically distinct $\pi$ and $\sigma$ orbitals whereas in the present case the two bands come from an on-molecule doublet. 

The total coupling $\lambda_{tot} = \sum_{ij}\lambda_{ij} N_i(0) /(N_i(0)+N_j(0)) = 2.29$, is exceptionally large, larger than other  found in other materials with strong electron-phonon couplings,\cite{Allen2000} but the combination of this $\lambda$ and our calculated density of states reproduces the measured specific heat. The value is consistent with that estimated by Andersen and collaborators \cite{Andersen1987} but inconsistent with other published estimates \cite{Fischer1978}.

\begin{figure*}[tb]
\includegraphics[width=1.0\columnwidth,clip]{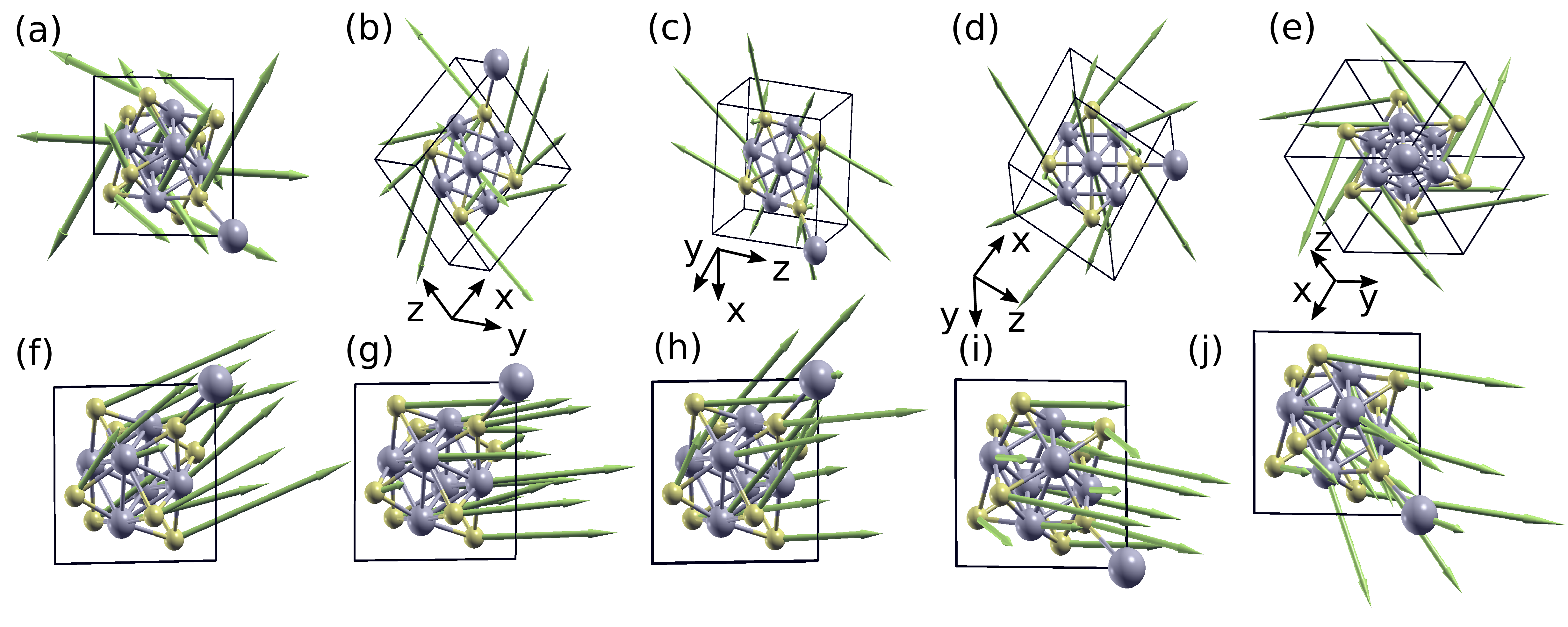}
\caption{The five $\gamma$-point phonon modes of PbMo$_6$S$_8$ with energies in the range of 11.0 meV to 15.6 meV ((a)-(e)) and the X point ((f)-(j)) of Brillouin zone, Pb atom is at one vertex of the cubic unit cell. Green arrows represent real space motion of atoms in the displayed phonon modes. Each sub-plot was generated by the {\it XCrySDen} package. \cite{Kokalj2003}} \label{phon_modes}
\end{figure*}

From Fig.~\ref{a2f}, one sees that the modes with largest coupling lie in the  frequency range from 11 meV to 17 meV. At the $\Gamma$ point, Pb-dominated modes form a low-lying transverse doublet around 5 meV and a longitudinal singlet around 10 meV. After these three modes, five modes can be observed below the gap at 17 meV. The atomic movement associated with these five phonon modes at the zone center ($\Gamma$ point) and zone boundary (X point) are represented in Fig.~\ref{phon_modes}. At the zone center, these five modes exhibit torsional character. External torsional modes have previously been suggested to be important for superconductivity based on a molecular crystal model. \cite{Bader1976_prl} At the zone boundary, these five modes show a character consistent with physics of dimerization, as a whole cluster rigidly moves towards its counterpart in the neighboring unit cell, albeit some mixing with other modes. Phonon modes with these characteristics are consistent with Peierls coupling.\cite{Mahan2000} At the zone center, phonons are limited to one unit cell, and rotations can impact the electron hopping between molecules by changing overlaps between molecular orbitals, since molecular orbitals are generally not spherical. (as shown in Fig.\ref{LUMO})  At the zone boundary, phonons are extended to two neighboring unit cells, and dimerization can modify electron hopping by changing the distance between molecules. Based on the above observations, we conclude that in Chverel phase compounds the most important contributions to the electron-phonon coupling are Peierls type couplings from 11 meV to 17 meV. 

\begin{figure}
\includegraphics[width=1.0\columnwidth,clip]{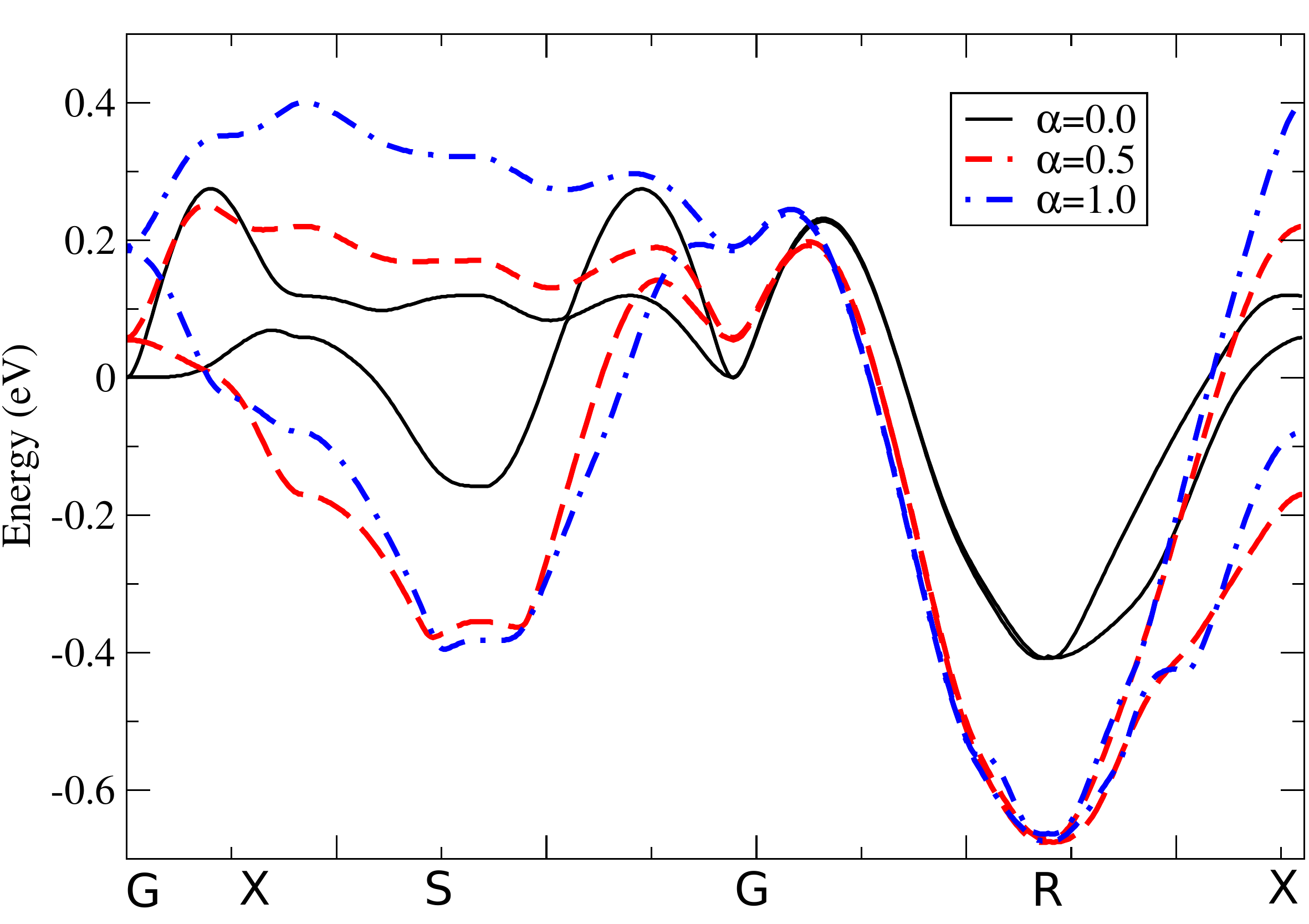}
\caption{ Band structure of the 11th phonon mode at the $\Gamma$ point, as shown in panel (e) of Fig.~\ref{phon_modes}. Atomic poitions {\bf X} for each calculation are determined by ${\bf X} = {\bf X_0} + \alpha {\bf u}$. {\bf X}$_0$ are the equilibrium atomic positions and {\bf u} is the phonon mode displacement from DFPT calculation.} \label{bs_phon}
\end{figure}

To further understand physics of those phonon modes, we calculated the variation of band structure due to the atomic displacement of the mode shown in panel (e) of Fig.~\ref{phon_modes}. As we can see in Fig.~\ref{bs_phon}, the band width increases with atomic displacement; but the degeneracy from the $\Gamma$ point to the R point is preserved. This degeneracy implies the phonon mode has no Jahn-Teller character; the increase of band width illustrates that the main effect is an increase in the overlap of each  Mo$_6$S$_8$ unit. This is the expected behavior from Peierls coupling: inter-molecular hoppings vary with vibrations; but intra-molecular states remain stationary. All information presented leads to the conclusion strongest electron-phonon coupling in Chevrel phase compounds occurs via Peierls active modes.   

\section{Consequences of The Electron-Phonon Interaction}\label{CEP}

\subsection{Normal State Self Energy}

The normal-state self-energy due to the electron-phonon interaction was calculated in the Migdal approximation, using the one-loop diagram with  non-interacting electron and phonon Green's functions and electron-phonon matrix elements obtained from our band structure. We separate the integral  over the electron momentum into an energy and a fermi surface integral and focussing on the band-diagonal terms in the self energy we obtain

\begin{widetext}
\begin{align}
\Sigma_{ii}(z)
  &= \int^{\infty}_{-\infty} d\epsilon \sum_{l\upsilon} \int^{\infty}_0 d\nu  \alpha^2F_{il}(\nu) [\frac{1+n_{\nu}^{\upsilon}-f(\epsilon)}{z - (\nu +\epsilon)}  - \frac{n_{\nu}^{\upsilon} + f(\epsilon)}{z-(-\nu + \epsilon)} \ ].
\end{align}

\end{widetext}

Here, {\it i,l} label the electronic bands, and $\upsilon$ labels phonon modes.

\begin{figure}
\includegraphics[width=1.0\columnwidth,clip]{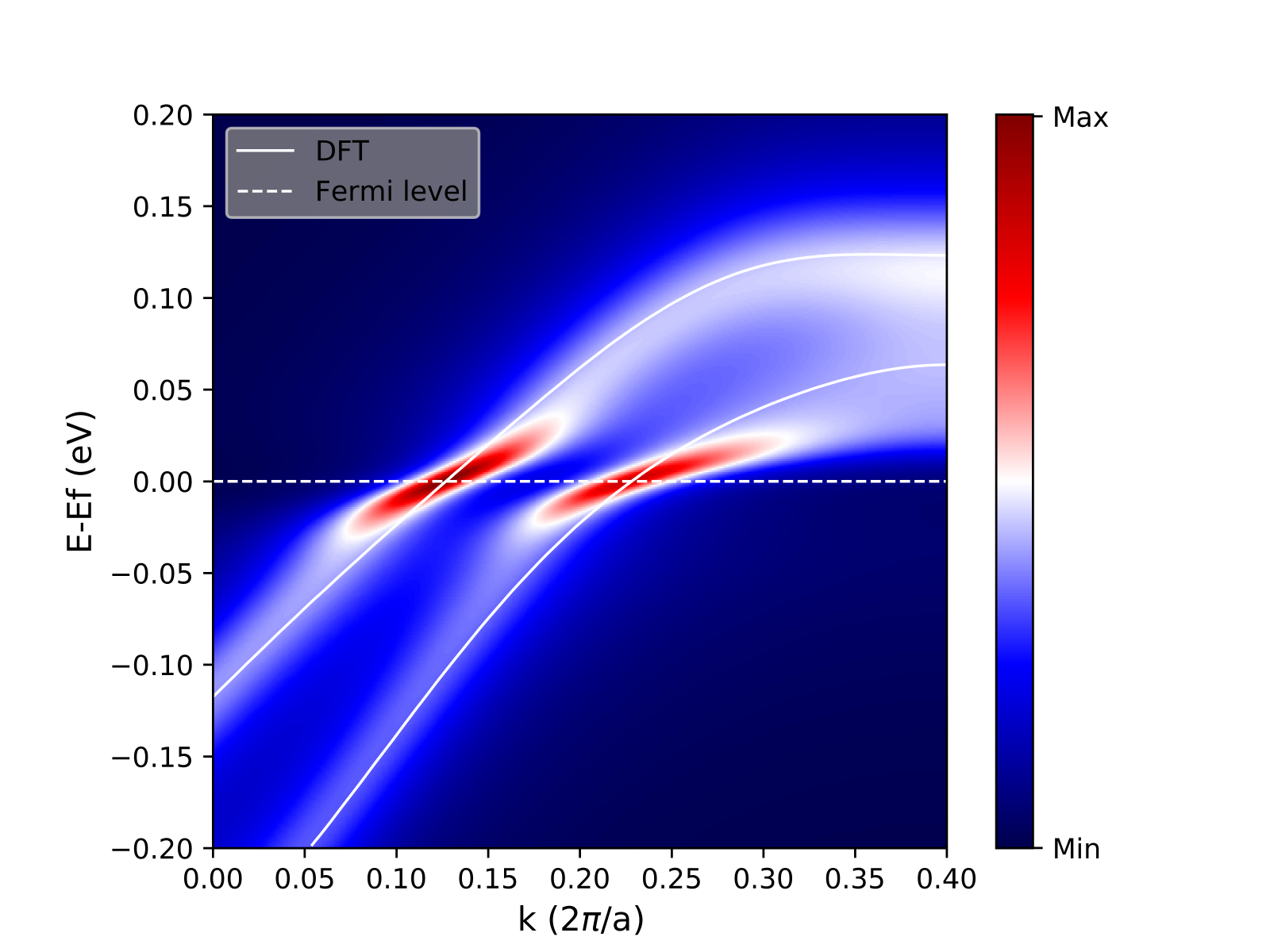}
\caption{False-color representation of electron spectral function with (shaded) and without (white line) electron-phonon interactions for near Fermi-surface momenta along the line from the  R to the X point of the Brillouin zone.} 
\label{ARPES}
\end{figure}

After analytically continuing the frequency argument $z$ to the real axis, we compute the electron spectral function at $T=0$ as
\begin{align}
 A_i(\textbf{k},\omega) = \frac{1}{\pi} \Im \left(\frac{1}{\omega - \epsilon_i(\textbf{k})-\Sigma_{ii}(\omega - i\delta)} \right).
\end{align}
Results are shown in  Fig.~\ref{ARPES}. We see that the electron-phonon interaction significantly modifies the dispersion only for energies within $\sim$ 20 meV of the fermi surface, leading to velocity renormalization of a factor of 2-3 at these energies. The near correspondence of bare and renormalized velocities at higher energies shows that  phonons at higher frequencies, including the internal Jahn Teller modes at $\sim$ 30 meV, have a relatively small effect on the spectrum.

\subsection{Superconductivity}\label{SC}

With band-resolved electron-phonon spectral function defined in Eq.~\ref{a2f_eq} and the self-energy evaluated in the Migdal approximation, we study strong coupling two-band superconductivity using the Eliashberg equations, following previous work on MgB$_2$\cite{Nicol2005} and  Mg$_{1-x}$Al$_x$B$_2$.\cite{UMMARINO2004} The equations may be written on the imaginary axis as

\begin{equation}\label{e1}
\begin{aligned}
& \Delta_i(i\omega_n)Z_i(i\omega_n) = \\ 
 &\pi T\sum_{m,j}[\lambda_{ij}(i\omega_m-i\omega_n)-\mu_{ij}^*]\frac{\Delta_j(i\omega_m)}{\sqrt{\omega_m^2+\Delta_j^2(i\omega_m)}} ,
\end{aligned}
\end{equation}

\begin{align}\label{e2}
Z_i(i\omega_n)=1+\frac{\pi T}{\omega_n}\sum_{m,j}\lambda_{ij}(i\omega_m-i\omega_n) \frac{\omega_m}{\sqrt{\omega_m^2+\Delta_j^2(i\omega_m)}} ,
\end{align}

where $\lambda_{ij}$ is:
\begin{align}
\lambda_{ij}(i\omega_m-i\omega_n)=2\int^{\infty}_0 d\Omega \frac{\Omega\alpha^2F_{ij}(\Omega)}{\Omega^2+(\omega_n-\omega_m)^2} .
\end{align}

We estimated the Coulomb pseudopotential $\mu^*_{ij}$  within this theory via $\mu_{ij} = U \sqrt{N_i(0)N_j(0)}$. $U=0.28$ eV used here is from a cRPA calculation, which gives $\mu_{ij}  \approx 1.4$. The Coulomb pesudopotential reduced by retardation effects leads to\cite{Morel1962}
\begin{align}\label{pseudopotential}
\mu_{ij}^*=\frac{\mu_{ij}}{1+\mu_{ij} \ln (E_{ele} / \omega_{ph})} .
\end{align}

The typical electron energy $E_{ele}$ is approximated by the half band width of {\it E$_g$} bands: $W / 2 \approx 0.35$ eV, and the relevant phonon frequency $\omega_{ph} \approx 12$ meV. (See the Bergmann-Rainer analysis below) This method yields a Coulomb pseudopotential value of $\mu_{ij}^* \approx 0.24$.

\begin{figure}
\includegraphics[width=1.0\columnwidth,clip]{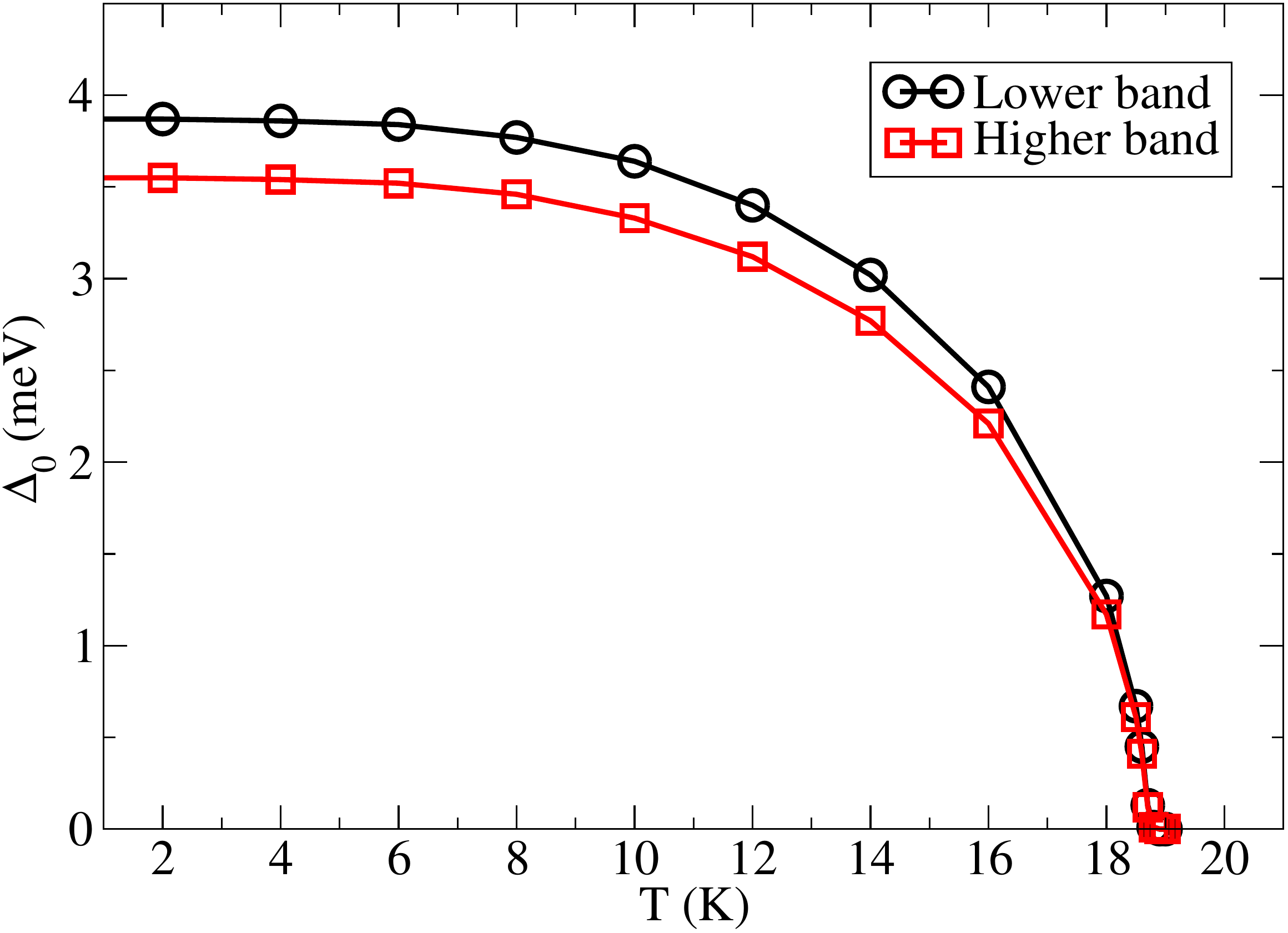}
\caption{Calculated superconducting gaps as function of temperature.} 
\label{Tc}
\end{figure}

\begin{figure}
\includegraphics[width=1.0\columnwidth,clip]{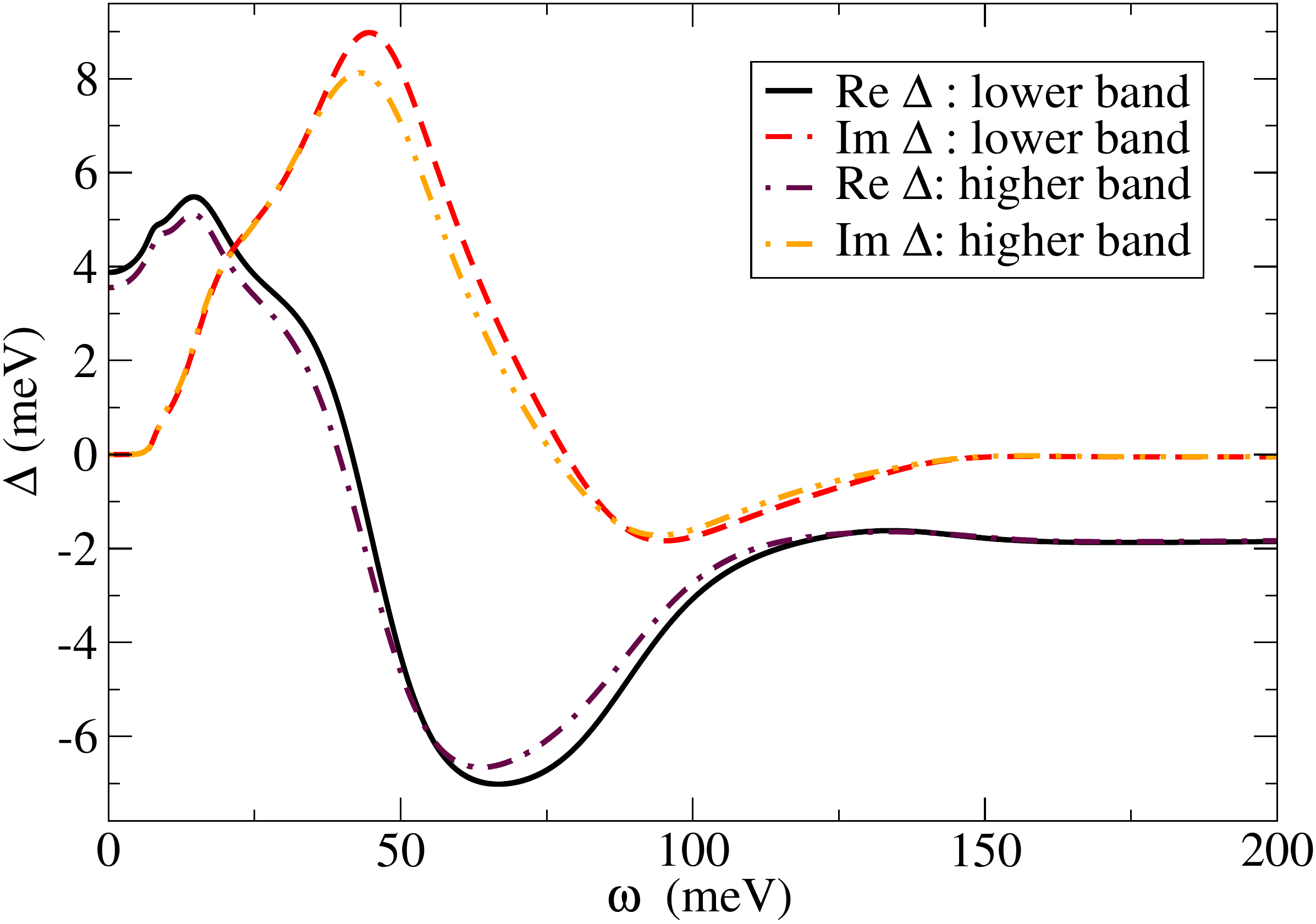}
\caption{Real and imaginary part of the two gap functions as a function of energy as extracted from Pad\'e approximants} 
\label{pade}
\end{figure}

Eq.~\ref{e1} and Eq.~\ref{e2} were solved on the imaginary axis and the gap functions were analytically continued to the real axis via Pad\'e approximants,\cite{Vidberg1977} as shown in Fig.~\ref{pade}. Superconducting gaps at the Fermi level as function of temperature are shown in Fig.~\ref{Tc}. The T$_c$ from our calculation is found out to be 18.8 K, which is larger than experimental value of 15 K by about 25\%. In the framework of the two-band isotropic Eliashberg equations used in this work, two possible reasons for this are the inadequate treatment of the Coulomb interaction and the anisotropy of Fermi surfaces. As shown in recent work, \cite{Bauer2012, Bauer2013} retardation effects are less effective in systems with strong coupling and narrow bands, which is the case for Chevrel phase compounds. In order to reproduce the experimental T$_c$ with the calculated $\alpha^2 F_{ij}$, the Coulomb pseudopotential would need to be $\mu_{ij}^* \approx 0.9$. We found two isotropic superconducting gaps are $\Delta_1 = 3.93$ meV and $\Delta_2 = 3.59$ meV. Scanning tunneling spectroscopy shows $\Delta_1 = 3.1$ meV and $\Delta_2 = 1.4$ meV.\cite{Petrovic2011} The large gap from our calculation is reasonable, but the overestimation of the smaller gap is significant. This discrepancy may arise from an exaggeration of $\alpha^2F$ from the DFPT calculations. Anisotropic calculations based on $\alpha^2F$ from DFPT also overestimate T$_c$ for multi-band superconductors such as MgB$_2$ \cite{Margine2013} and Ca-intercalated bilayer graphene. \cite{Margine2016}  

Earlier interest in Chevrel phase superconductors stemmed from their very high upper critical field $H_{c2}$, which can be related to coherence length $\xi_0$ via $H_{c2} \propto 1 / \xi_0^2$. Indeed a very  short coherence length (20\AA) has been reported based on magnetic measurements.\cite{Zheng1995} We can estimate coherence length within BCS theory via the superconducting gap and the Fermi velocity $\xi_0=\frac{\hbar v_F}{\pi \Delta}$. We calculated the Fermi velocities for two bands based on the DFT band structure, and they are renormalized by the electron-phonon coupling as $v_F^{*i} = v_F^i / (1+\sum_j\lambda_{ji})$. For the lower band $v_F^{*lb} = 1.09$ eV$\cdot$\AA, $\xi_0^{*lb} = 173$ \AA ~and the higher band $v_F^{*hb} = 0.85$ eV$\cdot$\AA, $\xi_0^{*hb} = 136$ \AA. The calculated coherence length is about one order of magnitude larger than the those reported in experiment. This is not necessarily a contradiction with experiment, Chevrel phase superconductors are known to be found in the dirty limit, \cite{Fischer1978, PENA201595} which implies measured coherence length is not an intrinsic property of pure crystal. Previously the mean-free path $l$ was estimated to be about 4~\AA. \cite{Fischer1976} $\xi=\sqrt{l \times \xi^*}$ gives a coherence length about 25~\AA, which is very close to reported experimental number 20~\AA.

We now extend the calculations to the other Chevrel phase compounds, assuming the electron-phonon matrix elements $g_{ij}^{\upsilon}(\textbf{k},\textbf{p})$ take on values of those of PbMo$_6$S$_8$, but using the material specific electronic band structures. Five Chevrel phase compounds were studied, and they can be put into two categories: M$^{2+}$Mo$_6$S$_8$ and M$^{3+}$Mo$_6$S$_8$, corresponding to two distinct doping levels for the Mo$_6$X$_8$ units. It is established that Yb, Sn and Pb belong to the first type and Y and La belong to the second type. \cite{PENA201595} As shown in Table ~\ref{chemical}, the main difference between those two types is the occupation of lower band around Fermi level. For M$^{2+}$Mo$_6$S$_8$, occupation of the lower band is incomplete, so there still is large DOS at the Fermi level. For M$^{3+}$Mo$_6$S$_8$, the occupation of the lower band is close to full and the occupation of higher band is close to half. As a result, the lower band has very little contribution to the DOS at the Fermi level, and one finds an effectively a single band situation.

As shown in Table ~\ref{chemical}, our calculations reproduce the experimental trends across material family very well. The lattice constants are quantitatively reproduced as is the variation of the transition temperatures. The calculated transition temperatures correlate with the total density of state at the Fermi level. Our calculation overestimates the absolute transition temperatures, with the  overestimation being larger for the lower T$_c$ values. A more detailed study of electron-phonon coupling across the entire material family is an important topic for future research.

\begin{table*}[t]
\caption{ Experimental and calculated values of T$_c$, lattice constants, and the DOS for five Chevrel phase compounds. 1 and 2 label the lower and the higer band. Experimental data are from Ref.~\protect\citen{PENA201595} and Ref.~\protect\citen{Hampshire2002}.   }
\label{chemical}
\begin{tabular}{l  c   c  c  c  c   c r }
\hline
   & Exp. Tc & Cal. Tc & Exp. {\it a} (\AA) &Cal. {\it a} (\AA)&N$_1$(0) & N$_2$(0)  & N(0)$_{tot}$ \\
 \hline
PbMo$_6$S$_8$ & 15.0 & 18.8 & 6.55 & 6.55 & 5.66 & 5.10 & 10.78 \\
SnMo$_6$S$_8$ & 13.0 &  17.2 & 6.52 & 6.52 &6.12 & 4.30 & 10.42\\
YbMo$_6$S$_8$ & 8.8 & 16.2 & 6.50 & 6.49 &  3.52  & 6.74 & 10.26 \\
LaMo$_6$S$_8$ & 7.1 & 11.0 & 6.51 & 6.52 & 0.94 & 6.60 & 7.54\\
YMo$_6$S$_8$  & 3.0 & 7.6 & 6.45 & 6.46  & 0.28 &  5.58 & 5.96 \\
 \hline
\end{tabular}
\end{table*}

To further address the question of which phonon modes are  most important for superconductivity, we calculate the functional derivative of T$_c$ with respect to $\alpha^2F_{ij}(\omega)$ following the scheme invented by  Bergmann and Rainer, \cite{Bergmann1973} and later extended to two-band systems by Mitrovi\'c.\cite{Mitrovic2004} The inter-band spectral functions are not independent: $\alpha^2F_{ij}(\omega) / \alpha^2F_{ji}(\omega) = N_j(0) / N_{i}(0)$. Only their combination as expressed through the off-diagonal spectral function defined in Eq.~\ref{off_diagonal} is meaningful,\cite{Dolgov2008} 

\begin{align}\label{off_diagonal}
\alpha^2F_{od}(\omega) = \frac{N_i(0)\alpha^2F_{ij}(\omega) + N_{j}(0)\alpha^2F_{ji}(\omega)}{N_i(0) + N_j(0)} .
\end{align}

\begin{figure}
\includegraphics[width=1.0\columnwidth,clip]{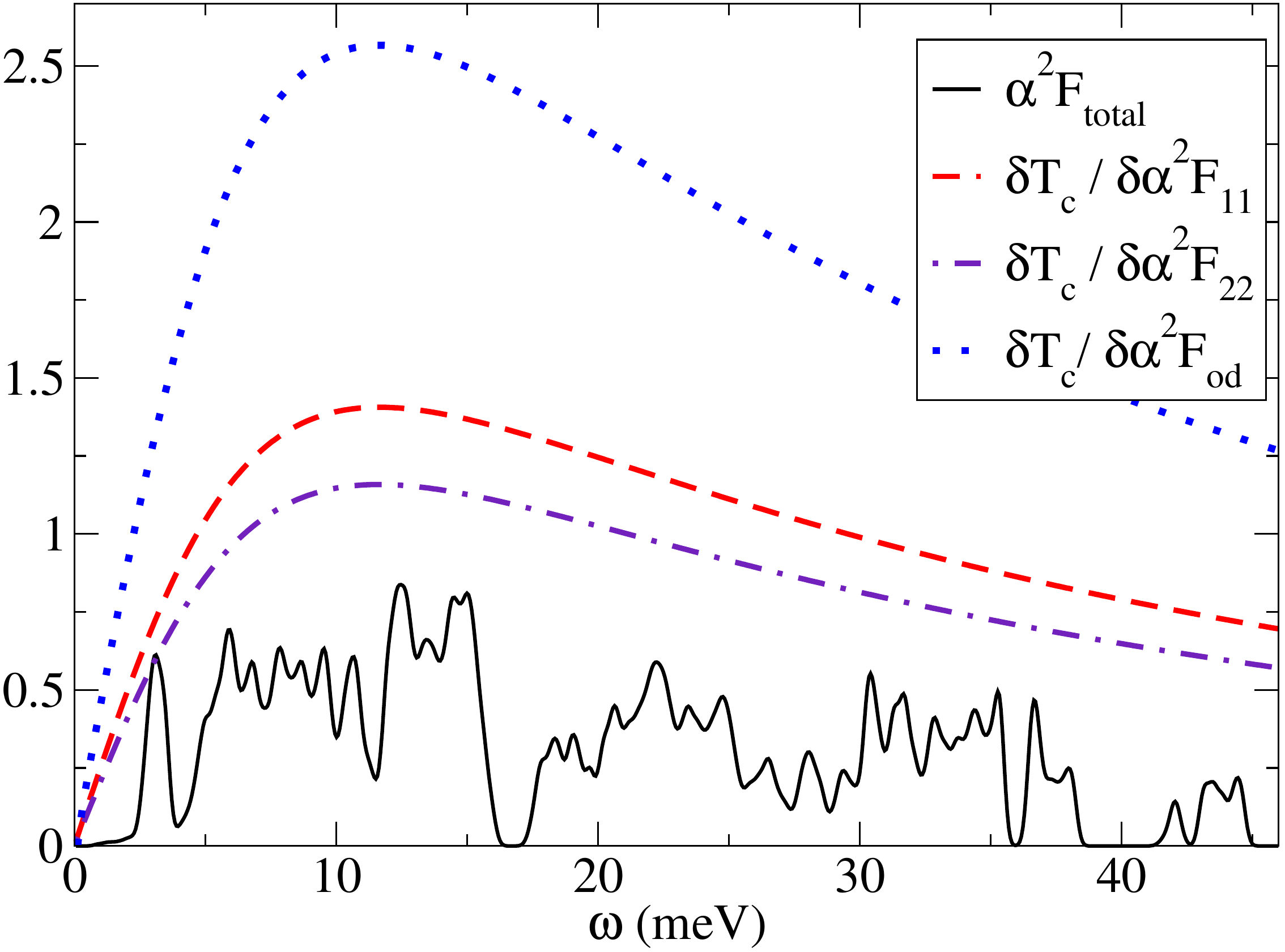}
\caption{Total electron-phonon spectral function, and functional derivative of T$_c$ with respect to band-resolved spectral function. 1 and 2 label lower and higher band} 
\label{derivative}
\end{figure}

Functional derivatives of relevant quantities are shown in Fig.~\ref{derivative}. Since three the $\alpha^2F_{ij}$ are not very different, it is expected that their functional derivatives show similar features. At low frequencies, the functional derivatives increase linearly with frequency, and they reach a maximum at about 12 meV. This number is close to earlier suggestions based on the comparison of low frequency phonons in PbMo$_6$S$_8$ and PbMo$_6$Se$_8$. \cite{Bader1976_prl, Schweiss1976, Bader1976}

As shown in section~\ref{isolated}, Jahn-Teller active intra-molecular modes are found at much higher frequencies than 12 meV. The Bergmann-Rainer analysis shows their relevance to superconductivity is eclipsed by modes at lower frequencies. Combined with the fact that phonons from 11 meV to 17 meV have the most important effect on the normal state spectrum, it is clear that phonon modes in this frequency range are the drivers of superconductivity in Chevrel phase compounds.

This finding is significant because  inter-molecular phonon modes are generally thought to not be relevant for superconductivity. As mentioned in section~\ref{intro}, superconductivity in faced-centered cubic X$_3$C$_{60}$ is thought to mainly arise from intra-molecular vibrational modes.\cite{Varma1991, Gunnarsson1997} On the other hand, Peierls couplings are frequently discussed in the context of metal-insulator transitions in low-dimensional materials. In particular, it has been shown for one-dimensional organic conductors, the Peierls instability suppresses superconductivity at lower temperatures.\cite{COLEMAN1973, Ferraris1973, Patton1973}  Our work shows that the Peierls coupling is important for superconductivity in 3D crystal such as Chevrel phase compounds.

\section{Conclusion}\label{conclusion}

We studied intra and inter-molecular interactions in Chevrel phase compounds, using PbMo$_6$S$_8$ as a model compound. Band structure calculations revealed two bands around the Fermi level which originate from two $E_g$ molecular orbitals and are about 0.7 eV wide. Constrained random phase approximation calculations estimated an on-site Hubbard U value of $U = 0.28$ eV and a value of Hund's exchange $J=0.04$ eV. Moreover, quantum chemistry calculations of isolated molecules were carried out to parameterize the Jahn-Teller effect in Mo$_6$X$_8$ molecules. The Jahn-Teller stability energy is $E_{JT} = 0.18$ meV, which is smaller than the band kinetic energy and intra-molecular Coulomb interaction values, but larger than the Hund's exchange. This energetic ordering is consistent with a metallic ground state.  If the band kinetic energy can be reduced via methods like chemical intercalation to the extent that materials are in the strongly correlated regime, the ground state could be a non-magnetic insulator because molecular Jahn-Teller effect suppresses Hund's coupling.\cite{Fabrizio1997} 

Density functional perturbation theory calculations with Wannier interpolations yield very strong electron-phonon coupling values, with $\lambda_{tot}=2.3$. Visible modifications to the electronic bands near the Fermi level can be found in our calculated ARPES spectra. Band-resolved electron-phonon spectral functions reveal that the largest couplings are due to phonon modes in frequency range from 11 meV to 17 meV. Phonon modes in this frequency range show the characteristics of Peierls-active modes.  

Superconductivity was studied by two-band Eliashberg equations, with band-resolved electron-phonon spectral functions. Superconducting properties, T$_c$ and the larger superconducting gap are all in reasonable agreement with experiments. Our current theory overestimates the the smaller superconducting gap. A Bergmann-Rainer analysis revealed that the most important phonon modes for superconductivity have frequencies around 12 meV, which is the spectral location of the largest electron-phonon coupling in PbMo$_6$S$_8$. To conclude,  our work showcases the importance of inter-molecular couplings for collective electronic behavior in molecular solids by illustrating an vital aspect that is overlooked in the standard molecular crystal model.\cite{HOLSTEIN1959}  Internal Jahn-Teller active modes which should be important for ground state magnetic properties in the strongly correlated regime, are not responsible for superconductivity in Chevrel phase compounds.

\section{Acknowledgements}
J.C. is supported by the NSF MRSEC program through Columbia in the Center for Precision Assembly of Superstratic and Superatomic Solids under Grant No.DMR-1420634. We uses resources of the National Energy Research Scientific Computing Center, a DOE Office of Science User Facility supported by the Office of Science of the U.S. Department of Energy under Contract No. DE-AC02-05CH11231.

\bibliography{Collect}

\end{document}